\documentclass[
12pt,
onecolumn,
 pra,
 amsmath,
 amssymb,
 longbibliography
]{revtex4-2}

\usepackage{graphicx}
\usepackage{dcolumn}
\usepackage{xcolor}
\usepackage{bm}
\usepackage{hyperref}

\usepackage{ulem}

\definecolor{apscite}{RGB}{0,120,160}

\usepackage[T1]{fontenc}
\usepackage{textcomp}

\hypersetup{
 colorlinks=true,
 citecolor=apscite,
 linkcolor=apscite,
 urlcolor=apscite
}

\begin{document}

\title{Energy-scalable single-burst attosecond emission at kiloelectronvolt photon energies} 

\preprint{260704 $\beta$1 version}

\author{Kaito Nishimiya}
\affiliation{Extreme Photonics Research Team, RIKEN Center for Advanced Photonics, RIKEN, 2-1 Hirosawa, Wako, Saitama 351-0198, Japan}

\author{Eiji J. Takahashi}
\email{ejtak@riken.jp}
\affiliation{Extreme Photonics Research Team, RIKEN Center for Advanced Photonics, RIKEN, 2-1 Hirosawa, Wako, Saitama 351-0198, Japan}
\date{\today}

\begin{abstract}
Attosecond soft-X-ray pulses at kiloelectronvolt photon energies would provide direct access to the L-edges of transition metals, enabling element-specific studies of charge, spin, and orbital dynamics on their intrinsic timescales. 
However, attosecond emission in this spectral region has remained elusive because the only phase-matched keV high-harmonic source demonstrated to date employed multicycle driving pulses without carrier-envelope-phase (CEP) control. A long-standing gap has therefore persisted between keV photon energies and CEP-controlled attosecond emission.
Here, we bridge this gap using a CEP-stabilized mid-infrared laser system with pulse durations tunable to the sub-cycle regime. Through phase-matched high-harmonic generation in low-pressure neutral helium under meter-scale loose-focusing conditions, we generate coherent soft-X-ray continua reaching 1.2~keV. Crucially, we observe CEP-dependent spectral modulation that reaches into the transition-metal L-edge region, providing experimental evidence consistent with the generation of isolated attosecond pulses. 
To demonstrate the capabilities of our source, we perform broadband soft-X-ray absorption spectroscopy spanning the Ti, Fe, Co, and Ni L-edges, as well as the O K-edge, and resolve near-edge fine structures. 
By combining meter-scale loose focusing with low-pressure phase matching, our approach addresses key energy-scaling limitations of previous keV high-harmonic sources and establishes a phase-matched, CEP-controlled keV soft-X-ray platform for attosecond spectroscopy of magnetic, quantum, and strongly correlated materials.
\end{abstract}

\keywords{attosecond pulse, high-order harmonics, XANES}

\maketitle

\section{Introduction}\label{sec1}

High-order harmonic generation (HHG) is the only demonstrated route to tabletop attosecond light sources, enabling time-resolved studies of electron dynamics on their intrinsic timescales across physics, chemistry, and materials science \cite{bib1,bib2,bib3}. 
Recent advances have established HHG-based spectroscopy \cite{bib48} as a powerful tool for probing ultrafast molecular dynamics through element-specific K-edge spectroscopy in the water-window region (284--540~eV) \cite{bib5,bib6,bib8} and through transition-metal M-edge spectroscopy in the extreme-ultraviolet (12.4--124~eV) region \cite{bib10,bib11}. In particular, attosecond spectroscopy has emerged in the water-window region \cite{bib4,bib7,bib9}, providing direct insight into ultrafast electronic dynamics in atoms and molecules. 
In contrast, HHG-based attosecond spectroscopy at the L$_{2,3}$ edges of 3d transition metals in the 700--900~eV soft X-ray (SXR, 124~eV--12.4~keV) region remains largely unexplored, although ultrafast SXR spectroscopy at these edges has previously been demonstrated \cite{bib50,bib51,bib53}. A laboratory-scale attosecond source covering this spectral range would provide direct access to the 2p--3d electronic transitions that underlie ultrafast magnetism, spintronics, quantum materials, and strongly correlated electron systems, enabling observation of light-induced demagnetization, spin transfer, and electron-correlation dynamics with subfemtosecond temporal resolution.

Achieving efficient HHG in the keV regime is fundamentally difficult. The harmonic cutoff scales favorably with the square of the driving wavelength ($\lambda$) \cite{bib13,bib14}; however, the single-atom conversion efficiency decreases approximately as $\lambda^{-5}$ \cite{bib15}. Furthermore, extending phase-matched HHG to keV photon energies generally requires increasingly demanding experimental conditions, such as very high gas pressures or intensity-modulation-based quasi-phase matching \cite{bib29,bib52}. Reconciling energy-scalable keV HHG with attosecond capability therefore remains an outstanding challenge. 
Realizing a practical attosecond platform at keV photon energies requires three conditions to be met simultaneously. First, the harmonics must be generated under phase-matched conditions that enable coherent buildup of the macroscopic emission and efficient scaling of the harmonic yield. Second, the driving pulse must be few-cycle and carrier-envelope-phase (CEP)-controlled, because isolated attosecond pulse (IAP) generation requires electron recollision to be confined to a single optical half-cycle. Third, the source architecture must support energy scaling of the generated SXR pulses while maintaining phase matching and CEP control, thereby enabling attosecond spectroscopy.
Previous studies have addressed these requirements only partially. CEP-dependent HHG under phase-matched conditions in neutral gases has been limited to the water-window region \cite{bib4,bib20,bib21,bib22}. In the 500--600~eV range, CEP-dependent emission has also been demonstrated using transient phase matching \cite{bib17} and overdriven propagation \cite{bib18}; however, these approaches rely on strong ionization, which introduces propagation distortions and thereby limits energy scalability \cite{bib19}. Furthermore, HHG schemes employing strongly ionized media have neither demonstrated CEP-dependent emission nor established a practical route toward energy-scalable keV attosecond sources \cite{bib28,bib47}. 

At higher photon energies, the only experimental demonstration of phase-matched HHG beyond 1~keV was reported by Popmintchev et al. in 2012, who used a tightly focused 3.9-$\mu$m laser pulse coupled into a helium-filled hollow waveguide at pressures approaching 40~atm \cite{bib29}. 
Although this landmark experiment established the feasibility of phase-matched keV HHG in a neutral medium, it employed multicycle driving pulses and therefore provided neither experimental evidence of CEP-dependent emission nor evidence consistent with IAP generation. 
Moreover, the extreme gas pressures and stringent waveguide requirements substantially increase the engineering complexity of the experimental configuration. 
Consequently, a long-standing gap remains between energy-scalable, phase-matched keV HHG and experimental evidence for IAP generation in this spectral region.

Here, we demonstrate a high-harmonic platform that simultaneously satisfies all three conditions for practical attosecond spectroscopy in the keV regime. 
Using a CEP-stable, 100-mJ-class mid-infrared laser system \cite{bib31} with pulse durations tunable from the few-cycle to the sub-cycle regime, we generate phase-matched high harmonics reaching 1.2~keV in a low-pressure neutral helium medium under a meter-scale loose-focusing geometry \cite{bib30}. 
We also observe clear CEP-dependent spectral intensity modulation at photon energies extending across the L-edges of ferromagnetic transition metals and the entire water-window region. 
These observations provide experimental evidence consistent with IAP generation, which has remained inaccessible to previously demonstrated multicycle keV HHG sources. To illustrate the practical utility of the source, we perform broadband SXR absorption spectroscopy spanning the Ti, Fe, Co, and Ni L-edges, as well as the O K-edge, in a single measurement, resolving characteristic near-edge spectral features across the entire spectral range. 
Finally, because phase matching is achieved in a low-pressure neutral medium without the extreme gas pressures and waveguide constraints required by previous keV HHG approaches, this platform provides a clear route toward energy scaling of the generated SXR pulses. 
Collectively, these results establish an energy-scalable, phase-matched, CEP-controlled keV SXR platform and open a pathway toward attosecond spectroscopy at transition-metal L-edges.

\section{Results}\label{sec2}

\subsection{Energy scaling strategy for keV attosecond emission}\label{subsec:energy-scaling}

The landmark demonstration by Popmintchev et al. of phase-matched HHG extending to 1.6~keV relied on tight focusing and helium pressures approaching 40~atm \cite{bib29}. 
Although this experiment established the feasibility of neutral-gas phase matching in the keV regime, it also highlighted a central challenge: phase matching at keV photon energies generally requires increasingly high gas pressures when tightly focused, long-wavelength drivers are employed.
Under free-space focusing, phase matching ($\Delta k = k_q - qk_0$) is governed by the balance among neutral-atom dispersion, plasma dispersion induced by laser ionization, and the Gouy phase. 
When the ionization fraction ($\eta$) exceeds the critical ionization fraction ($\eta_{\mathrm{c}}$, see Methods), plasma dispersion becomes too large to be compensated by neutral-atom dispersion, and the phase-matching condition can no longer be satisfied \cite{criticalI}. 
For a fixed pulse duration, the weak-ionization criterion therefore determines the maximum allowable peak intensity of the driving pulse.
To ensure robust phase matching in a weakly ionized neutral medium, we adopt the conservative condition $\eta = \eta_{\mathrm{c}}/2$. Under this condition, a practical route to overcoming the high-pressure requirements of phase-matched keV HHG with free-space focusing is provided by a loose-focusing scaling strategy \cite{bib30}.

\begin{figure}
\centering
\includegraphics[width=\textwidth]{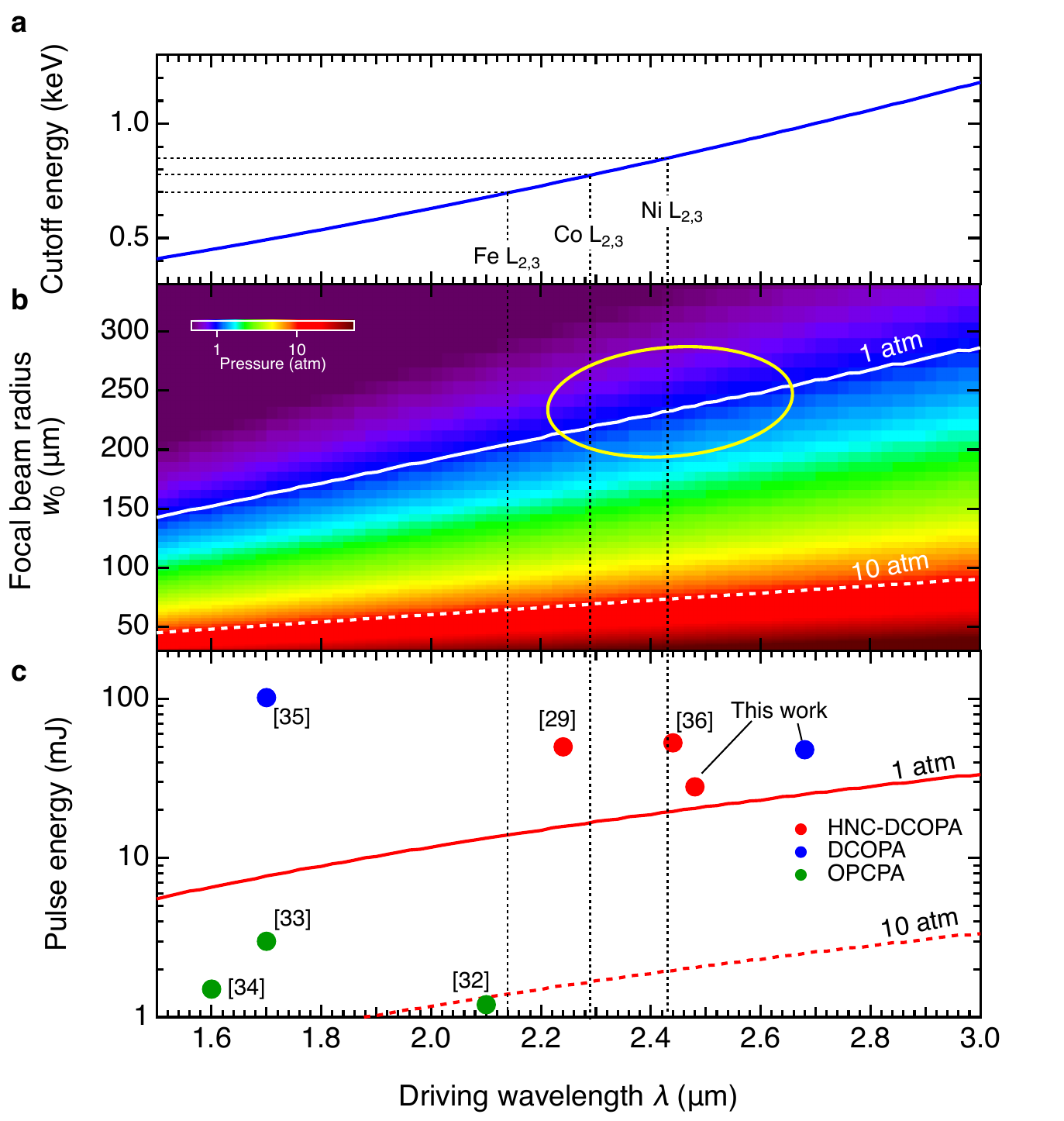}
\caption{
\textbf{Phase-matching conditions for keV HHG in a neutral gas.}\\
\textbf{a}, Calculated cutoff energy as a function of the driving wavelength under the weak-ionization condition ($\eta = \eta_{\mathrm{c}}/2$) for three-cycle driving pulses.
\textbf{b}, Calculated phase-matching pressure as a function of the driving wavelength and focal beam radius under the corresponding cutoff-energy condition. The yellow circle indicates the target operating region.
\textbf{c}, Calculated driving-pulse energy required to satisfy the weak-ionization condition ($\eta = \eta_{\mathrm{c}}/2$) along the 1-atm (solid line) and 10-atm (dashed line) phase-matching contours (solid and dashed white lines, respectively, in \textbf{b}). 
The plotted symbols represent CEP-stable laser systems with pulse durations of three optical cycles or less and pulse energies greater than 1~mJ \cite{CEPOPA1,CEPOPA2,CEPOPA3,CEPOPA4,CEPOPA5,xu_NP,bib31}.
}
\label{fig1}
\end{figure}

The weak-ionization condition, $\eta = \eta_{\mathrm{c}}/2$, determines the maximum allowable peak intensity of the incident driving pulse for a fixed pulse duration. 
Once the driving wavelength is specified, this maximum intensity determines the corresponding cutoff energy, as shown in Fig.~\ref{fig1}a. 
In this calculation, the pulse duration was set to three optical cycles at each wavelength to represent a CEP-controlled, few-cycle driving condition relevant to attosecond emission.
At sufficiently high harmonic orders, the optimum phase-matching pressure can be approximated as $p_{\mathrm{opt}} = \lambda^2/(\pi^2 \delta n w_0^2)$, where $\delta n$ is the refractive-index difference of the neutral gas at 1~atm and $w_0$ is the focal beam radius (Methods). 
Figure~\ref{fig1}b shows the calculated phase-matching pressure in helium as a function of the driving wavelength and focal beam radius. 
In a tight-focusing geometry with a focal beam radius of approximately 50~$\mu$m, the phase-matching pressure reaches several tens of atmospheres in the keV regime. 
Such high pressures not only impose a substantial gas load on the vacuum system but also result in impractical helium consumption, thereby complicating routine operation and scalability. 
This constraint has been a major obstacle to the scalability of keV HHG. 
Because the phase-matching pressure scales as $1/w_0^2$, increasing the focal beam radius to 200~$\mu$m reduces the required helium pressure to approximately 1~atm, corresponding to the target operating region indicated by the yellow circle in Fig.~\ref{fig1}b.
Figure~\ref{fig1}c shows the pulse energy required to reach the 1-atm and 10-atm phase-matching contours, indicated in Fig.~\ref{fig1}b by the solid and dashed white lines, respectively. This energy corresponds to increasing the focal beam radius while maintaining the peak intensity at $\eta = \eta_{\mathrm{c}}/2$ at each driving wavelength. 
This analysis shows that driving pulses of several tens of mJ are required to reach the keV regime. Under these conditions, the keV harmonic yield is expected to increase by approximately an order of magnitude relative to that obtained with the tight-focusing geometry, owing to an approximately 10-fold increase in beam cross-sectional area, while maintaining a practical ($\sim$1~atm) phase-matching pressure.
The plotted symbols represent CEP-stable laser systems with pulse durations of three optical cycles or less and pulse energies greater than 1~mJ \cite{CEPOPA1,CEPOPA2,CEPOPA3,CEPOPA4,CEPOPA5,xu_NP,bib31}. 
Among the systems shown in Fig.~\ref{fig1}c, our dual-chirped optical parametric amplification (DC-OPA) and heterogeneous-nonlinear-crystal DC-OPA (HNC-DCOPA) platforms are positioned in the regime required to achieve phase matching at pressures below 1~atm.
Further increases in pulse energy would allow proportionally larger focal beam sizes, thereby further reducing the required gas pressure and increasing the HHG yield. 
Loose focusing therefore shifts the primary scaling challenge from extreme gas pressure to high-pulse-energy laser systems, making phase-matched keV HHG accessible with modern mid-infrared drivers delivering pulse energies of several tens of mJ.
This scaling strategy also highlights that increasing the repetition rate alone cannot alleviate the stringent phase-matching pressure requirements for keV HHG because the phase-matching condition is fundamentally determined by the driving-pulse energy and focusing geometry rather than by the average power. 
Laser pulse energy, rather than average power, is therefore the key scaling parameter for reducing the phase-matching pressure in practical keV HHG.

\subsection{Phase-matched HHG reaching 1.2~keV with meter-scale loose focusing}\label{subsec:phase-matched-1p2kev}
\begin{figure}[hb]
\centering
\includegraphics[width=\linewidth]{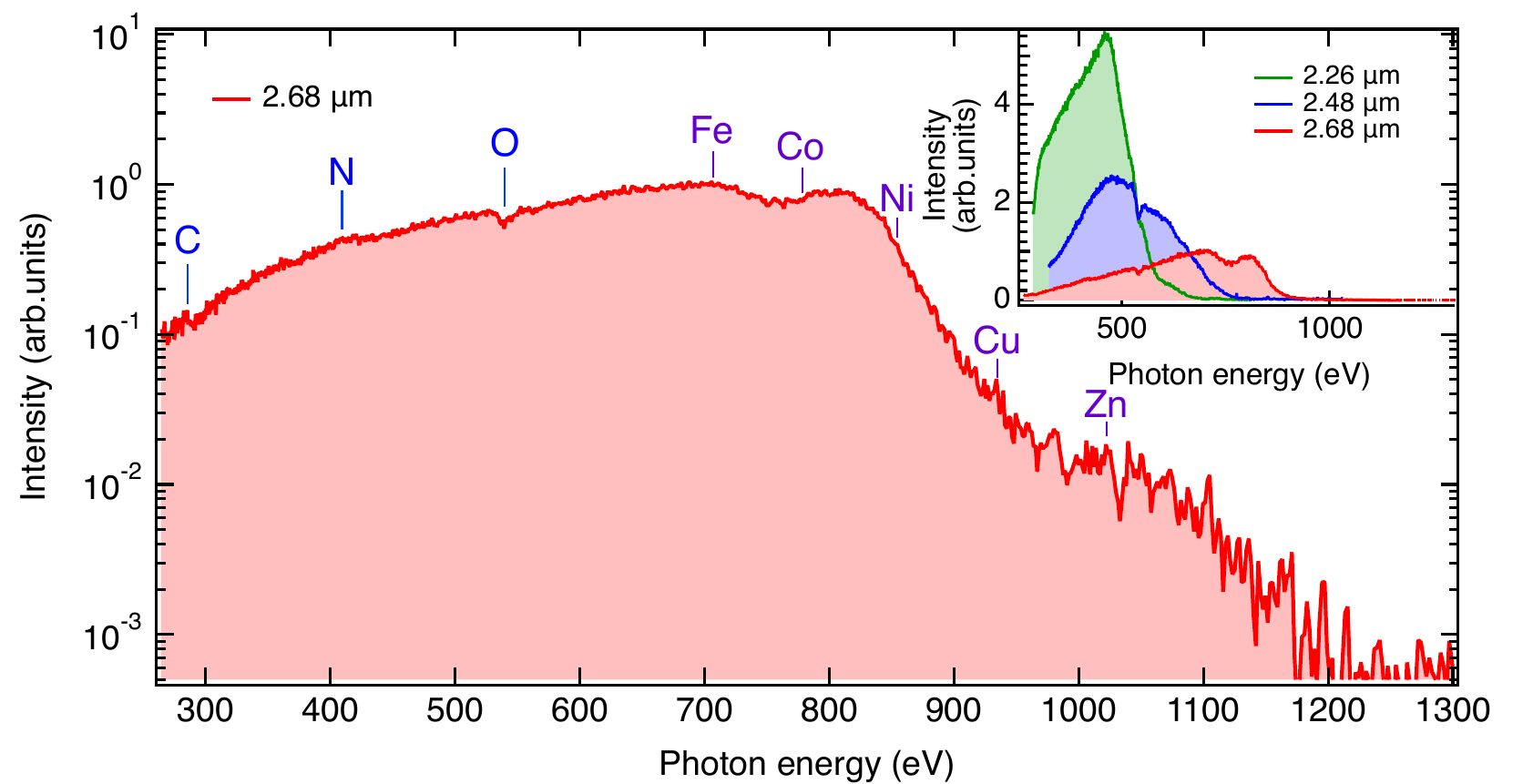}
\caption{
\textbf{Driving-wavelength dependence of high-order harmonic spectra.}\\
High-order harmonic spectrum generated with a driving wavelength of 2.68~$\mu$m, shown on a logarithmic scale.
The inset shows the dependence of the high-order harmonic spectra on the driving wavelength, plotted on a linear scale.
 All spectra were acquired with an accumulation time of 3000~s. 
}
\label{fig2}
\end{figure}

Guided by the energy-scaling analysis for keV HHG described above, we employed a CEP-stable high-energy mid-infrared laser system based on DC-OPA \cite{bib31,bib32}. 
The laser was focused using a concave mirror with a focal length of 3~m into a 50-mm-long gas cell. 
Details of the driving laser and HHG setup are provided in the Methods section. 
Figure~\ref{fig2} shows a high-harmonic spectrum generated in He using a driving laser wavelength of 2.68~$\mu$m, with the cutoff energy reaching 1.2~keV.
In these measurements, the CEP was left free-running, and the harmonic yield increased quadratically with the He gas pressure. 
This quadratic pressure dependence confirms coherent buildup of high-order harmonics under phase-matched conditions. 
The He gas pressure in the interaction region was reduced to approximately 1~atm, more than an order of magnitude below the tens of atmospheres required for phase matching in the keV region under conventional tight-focusing conditions. 
 In addition, the gas consumption during the experiment was reduced to approximately 2~L/min, substantially reducing the gas load on the vacuum system. 
 This low-pressure, low-gas-consumption operation is particularly advantageous for experiments requiring long-term signal accumulation, including attosecond spectroscopy and broadband SXR absorption measurements.
The inset shows the evolution of the high-order harmonic spectrum as the driving laser wavelength increases.
All spectra were acquired with an accumulation time of 3000~s.
As the driving wavelength increased, the cutoff extended to higher photon energies, whereas the high-order harmonic yield decreased.
The pulse energies above the carbon K-edge (284~eV), estimated at the generation point from the measured high-order harmonic spectra, were 1, 0.35, and 0.08~pJ for driving wavelengths of 2.26, 2.48, and 2.68~$\mu$m, respectively.
A direct quantitative comparison is not straightforward because the driving-pulse energies and durations, as well as the gas pressures, differed among the measurements. Nevertheless, the observed reduction in harmonic yield is stronger than that expected from the wavelength scaling of the single-atom response \cite{bib15}. 
These results indicate that the measured wavelength dependence cannot be attributed solely to the microscopic single-atom response and that macroscopic effects, including phase matching and group-velocity matching, must also be considered \cite{PM_rev,GVM}.
This ultra-broad spectral range directly spans two key spectroscopic regions: the water window, which contains the K-edges of light elements important in biological and chemical systems, and the transition-metal L-edge region, which encompasses the Fe, Co, and Ni L-edges central to magnetic studies. 
To be practical for attosecond spectroscopy, the source must also retain its attosecond capability under these phase-matched conditions. 
To assess whether CEP-controlled attosecond emission can be sustained under phase-matched conditions across the keV spectral range, we next examine CEP-dependent spectral modulation and pressure-dependent coherent buildup in both the transition-metal L-edge and water-window regions.

\subsection{CEP-dependent phase-matched HHG in the transition-metal L-edge region}\label{subsec:cep-l-edge}

\begin{figure}
\centering
\includegraphics[width=\linewidth]{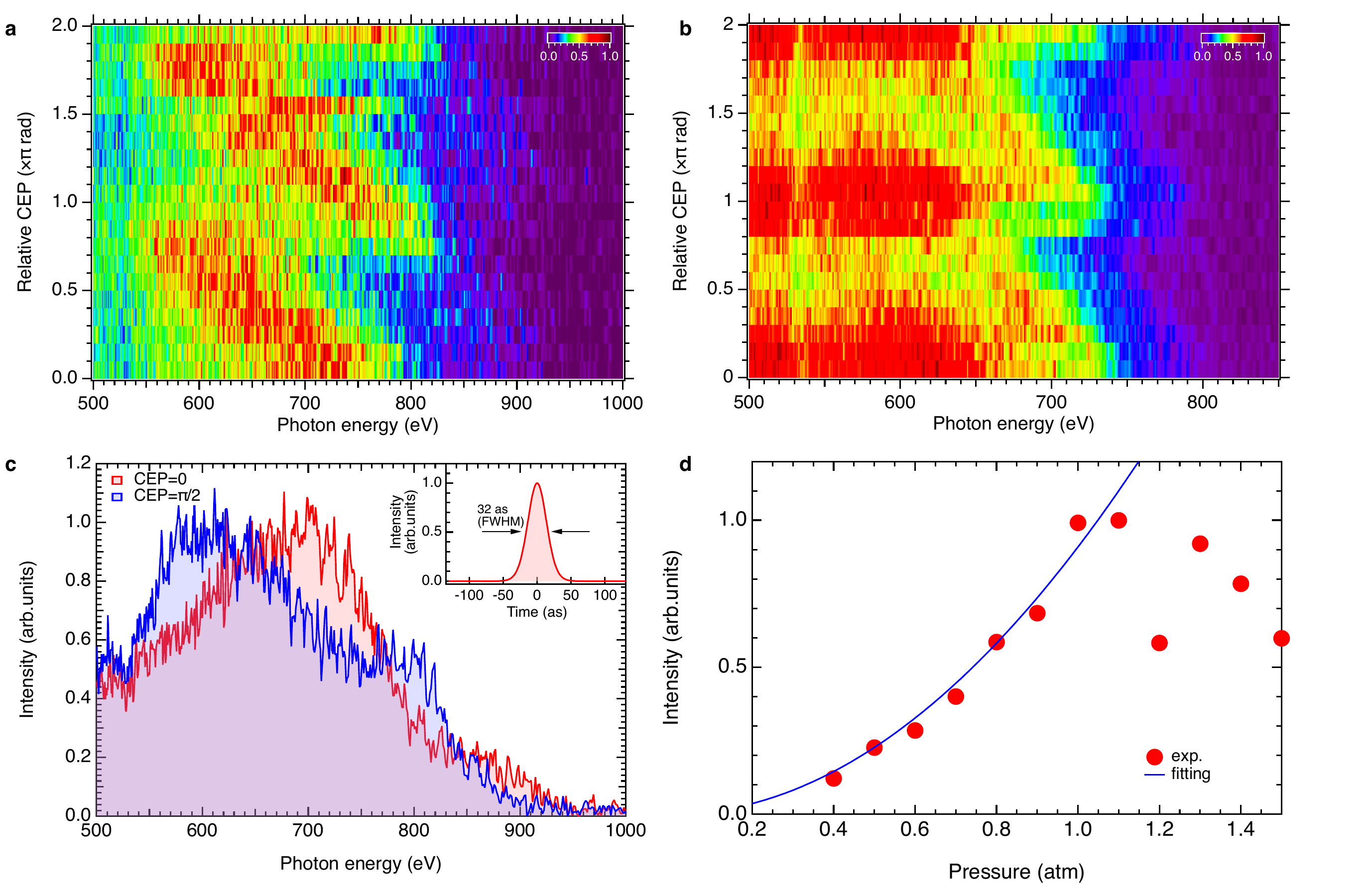}
\caption{
\textbf{CEP-controlled, phase-matched HHG in the transition-metal L-edge region.}\\
\textbf{a},\textbf{b}, CEP dependence of the high-harmonic spectra generated using (\textbf{a}) 2.68-$\mu$m and (\textbf{b}) 2.48-$\mu$m driving pulses.
\textbf{c}, Representative spectra extracted from the CEP scan shown in \textbf{a} at relative CEPs = 0 rad and $\pi/2$ rad. 
The upper-right inset shows the FTL pulse duration estimated from the continuum spectrum.
\textbf{d}, Pressure dependence of the harmonic signal at 850~eV.
}
\label{fig3}
\end{figure}

We first examine the transition-metal L-edge region, where the combination of CEP control and phase matching has not been demonstrated previously. 
Figure~\ref{fig3}a,b show the CEP dependence of the high-harmonic spectra obtained in this region using CEP-stable driving pulses at 2.68 and 2.48~$\mu$m. 
The laser parameters and HHG conditions are summarized in Supplementary Table~1. 
For the 2.68-$\mu$m driver, the harmonic cutoff extends to approximately 950~eV and exhibits pronounced CEP-dependent spectral intensity modulation. 
Figure~\ref{fig3}c compares spectra recorded at relative CEPs = 0 and $\pi/2$, revealing a cutoff shift of approximately 100~eV. 
A similar CEP dependence is observed for the 2.48-$\mu$m driver, for which a continuous spectral region spanning approximately 150~eV is obtained around 800~eV. 
Such CEP-dependent continuum formation provides experimental evidence consistent with single-burst emission \cite{bib_CEPIAP3,bib20,bib_CEPIAP1,bib_CEPIAP2,bib18,bib_CEPIAP4,bib17,bib_CEPIAP5}. 
Assuming a flat spectral phase, the Fourier-transform-limited (FTL) pulse duration estimated from the continuum near 900~eV is 32~as.

To determine whether the harmonic emission is generated under phase-matched conditions in a neutral medium, we measured the pressure dependence of the harmonic signal at 850~eV (Fig.~\ref{fig3}d). 
In the low-pressure regime, the harmonic intensity exhibits a clear quadratic dependence on gas pressure, consistent with coherent buildup of the harmonic field under phase-matched conditions. 
The peak intensity estimated from the cutoff energy is $4.5 \times 10^{14}$~W/cm$^2$, corresponding to a calculated phase-matching pressure of approximately 1.3~atm. 
This value agrees well with the experimentally observed optimum pressure. 
Together, the CEP-dependent spectral modulation and pressure-dependent coherent buildup provide experimental evidence for single-burst emission under phase-matched conditions in the transition-metal L-edge region.


\subsection{Phase-matched CEP-controlled supercontinuum spanning the entire water-window region}\label{subsec:water-window}

\begin{figure}[h]
\centering
\includegraphics[width=\linewidth]{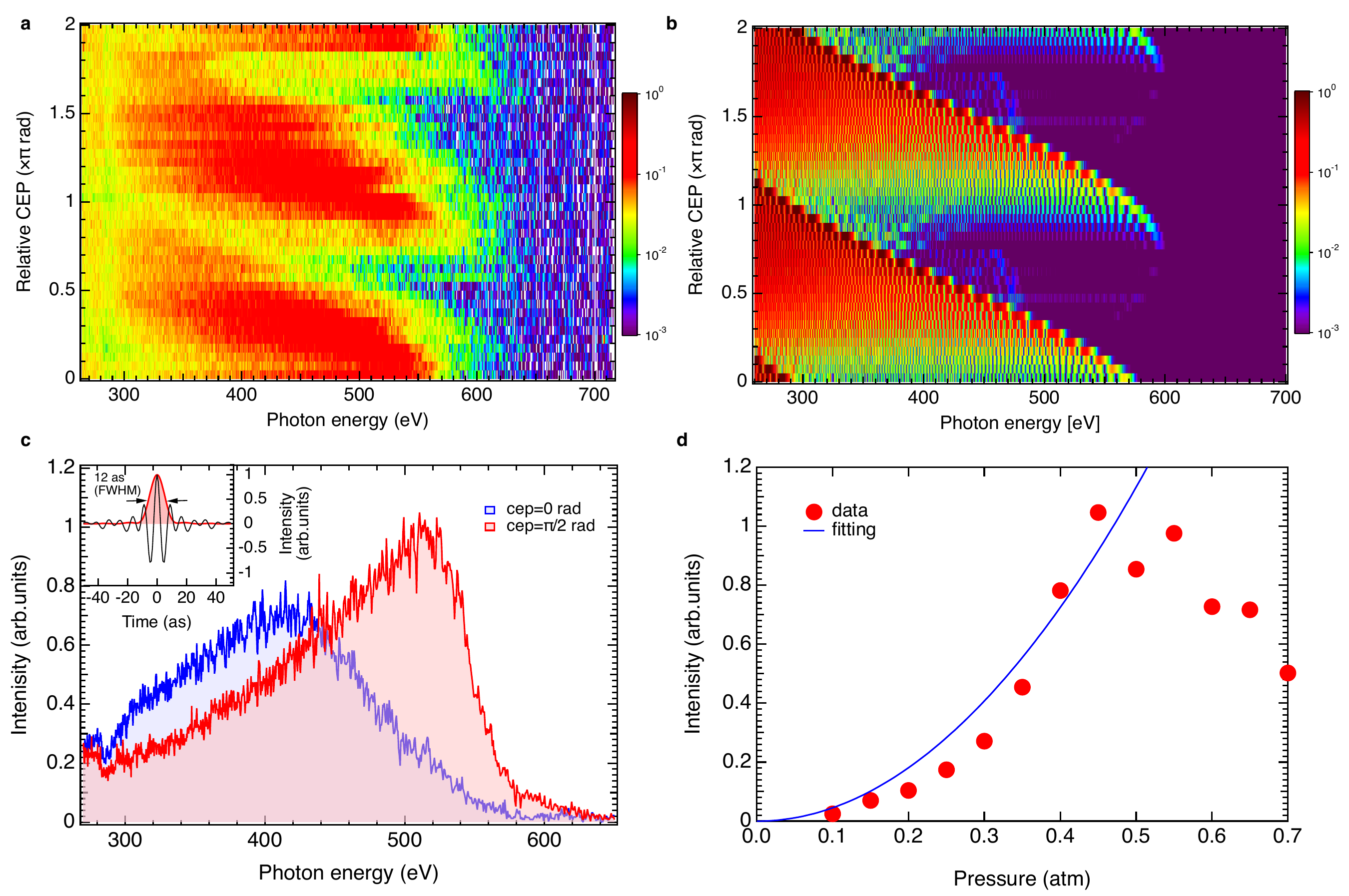}
\caption{
\textbf{CEP-controlled supercontinuum spanning the entire water window driven by a sub-cycle laser.}\\
\textbf{a}--\textbf{c}, CEP dependence of high-harmonic spectra.
\textbf{a}, Experimental results.
\textbf{b}, 1D-TDSE calculation results.
\textbf{c}, Representative line spectra extracted from the CEP scan shown in \textbf{a}. 
The inset shows the FTL pulse duration estimated from the continuum spectral region.
\textbf{d}, Pressure dependence of the harmonic signal at 450~eV. 
}
\label{fig4}
\end{figure}

The interpretation of the CEP-dependent emission observed in the transition-metal L-edge region can be further validated by measurements in the water-window region.
Our laser system provides sub-cycle driving pulses, for which the CEP dependence manifests as a pronounced intensity modulation extending over several hundred electronvolts. 
This operating condition shifts the central wavelength of the driver to shorter values, enabling the harmonic emission to cover the entire water-window region. Figure~\ref{fig4}a shows the CEP dependence of the high-harmonic spectrum generated using a sub-cycle 2.26-$\mu$m driver at a helium pressure of 0.6~atm. 
The cutoff energy extends to 620~eV, and the CEP-dependent continuum spans more than 300~eV, covering the entire water-window region. 
These experimental observations are supported by one-dimensional time-dependent Schrödinger equation (1D-TDSE) calculations (Fig.~\ref{fig4}b) \cite{TDSE}, which accurately reproduce the CEP-dependent cutoff shift. 
This agreement supports the interpretation that the observed continuum originates from CEP-controlled single-burst emission. 
Figure~\ref{fig4}c compares spectra recorded at relative CEPs = 0 and $\pi/2$, highlighting the pronounced CEP dependence and broadband continuum generation. 
Assuming a flat spectral phase, this continuum of more than 300~eV corresponds to an FTL pulse duration of 12~as (Fig.~\ref{fig4}c inset), shorter than one atomic unit of time (24~as).
To the best of our knowledge, this represents the first demonstration of a CEP-controlled supercontinuum spanning the entire water-window region, including the Ti L-edge, within a single spectrum. 
The resulting source therefore provides a platform for broadband, element-specific attosecond spectroscopy.

To examine the phase-matching conditions, we measured the pressure dependence of the harmonic signal at 450~eV (Fig.~\ref{fig4}d). 
At low pressures, the harmonic intensity exhibits a clear quadratic dependence on pressure, consistent with coherent buildup under phase-matched conditions in a neutral medium. 
The signal reaches a maximum near 0.6~atm and subsequently decreases at higher pressures, in good agreement with the phase-matching pressure predicted for these HHG conditions. 
Together, the CEP-dependent continuum generation and pressure-dependent coherent buildup demonstrate that the phase-matched, CEP-controlled HHG platform remains scalable across the entire water-window region.

\subsection{Broadband XANES spectroscopy at transition-metal L-edges}\label{subsec:xanes-l-edge}
Having established CEP-controlled HHG extending from the water window to the transition-metal L-edge region, we next demonstrate the capability of this source for broadband SXR absorption spectroscopy. 
X-ray absorption near-edge structure (XANES) measurements were performed at the L-edges of Ti, Fe, Co, and Ni, confirming that the ability of the source to provide broadband, element-specific SXR spectra is well suited to future attosecond absorption spectroscopy of transition metals. 
Access to these L-edges is essential for probing the electronic, orbital, and spin states that govern ultrafast magnetism and spintronic phenomena.

\begin{figure}
\centering
\includegraphics[width=1\linewidth]{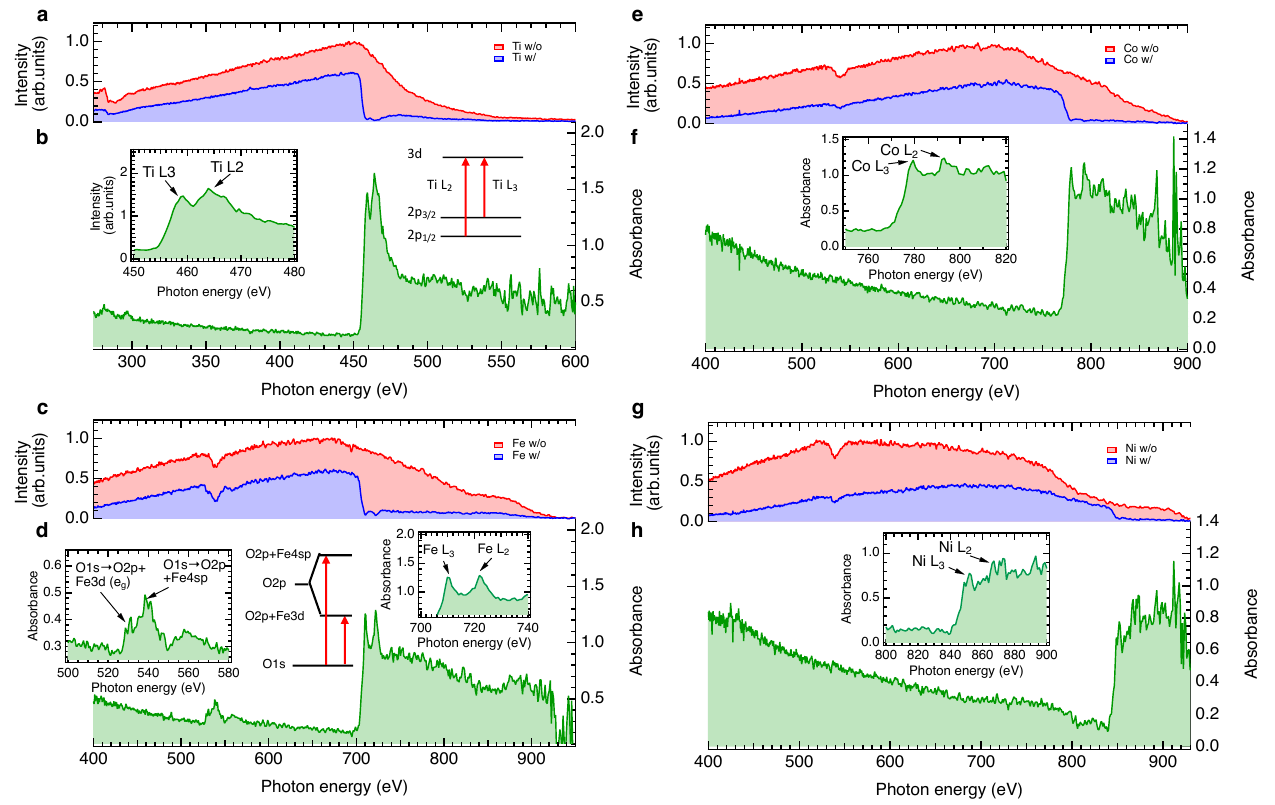}
\caption{
\textbf{Broadband transition-metal L-edge XANES using CEP-controlled SXR high harmonics.}\\
\textbf{a},\textbf{c},\textbf{e},\textbf{g}, HHG spectra measured without (red) and with (blue) 200-nm-thick Ti, Fe, Co, and Ni filters, respectively. 
The integration time and driver wavelength were 1000~s and 2.26~$\mu$m for Ti, and 4000~s and 2.68~$\mu$m for Fe, Co, and Ni. The CEP of the driving laser remained stable throughout the acquisition.
\textbf{b},\textbf{d},\textbf{f},\textbf{h}, Absorbance spectra of the Ti, Fe, Co, and Ni filters obtained from the corresponding reference and transmitted HHG spectra. 
The spin--orbit-split L$_3$ and L$_2$ absorption edges are clearly resolved for Ti, Fe, Co, and Ni. 
The insets show enlarged views of the respective L-edge regions. 
For Fe, additional O K-edge absorption features are observed because of surface oxidation of the filter.
}
\label{fig5}
\end{figure}

Figure~\ref{fig5} presents the absorption spectra of 200-nm-thick Ti (\textbf{a},\textbf{b}), Fe (\textbf{c},\textbf{d}), Co (\textbf{e},\textbf{f}), and Ni (\textbf{g},\textbf{h}) filters. The absorbance spectra were calculated as $-\log(I_{\mathrm{filter}}/I_0)$, where $I_{\mathrm{filter}}$ and $I_0$ represent the high-harmonic spectra measured with and without the metal filter, respectively. 
The Ti L-edge was measured using the broadband continuum generated with the CEP-stabilized, sub-cycle 2.26-$\mu$m laser, whereas the Fe, Co, and Ni L-edges were measured using broadband high-harmonic continua generated with the 2.68-$\mu$m laser.
In both measurements, the absorption spectra were acquired while the relative CEP of the driving laser was held constant. 
Together with the CEP-dependent continuum formation established above, this CEP stability enables the use of the cutoff region associated with single-burst emission for future attosecond time-resolved absorption spectroscopy.
For the Ti, Fe, Co, and Ni filters, the L$_{2,3}$ splitting characteristic of transition-metal L-edges was clearly resolved (Fig.~\ref{fig5}, insets). 
The absorption features observed in our spectra agree well with those measured by synchrotron spectroscopy \cite{bib33,bib34}.

Furthermore, multiple absorption structures corresponding to the O K-edge were observed for the Fe sample (Fig.~\ref{fig5}d). 
In the raw high-harmonic spectra measured with and without the Fe filter (Fig.~\ref{fig5}c), an oxygen absorption feature is observed near 540~eV, likely arising from contamination in the vacuum chamber or from the oxidized aluminum filter used to remove the fundamental beam.
In the absorbance spectrum of the Fe sample, however, this impurity-related absorption is canceled as a common background contribution, revealing multiple absorption features attributed to iron oxide and/or oxyhydroxide species formed by oxidation of the Fe filter in air. 
These features originate from hybridized O 2p--Fe 3d and O 2p--Fe 4sp states and agree well with those reported in synchrotron XANES spectra of iron oxides and oxyhydroxides \cite{bib35,bib36,bib37}. 
An oxygen-related absorption feature is clearly observed for Fe, whereas no corresponding feature is resolved for Co or Ni. 
Although Fe, Co, and Ni readily oxidize upon exposure to ambient air, Co and Ni form approximately nanometer-thick passivating oxide layers that suppress further oxidation \cite{ioni_co,ioni_ni}.
In contrast, the hydrated oxide formed on Fe provides little passivation, allowing oxidation to proceed further into the material and resulting in the heterogeneous growth of oxide/oxyhydroxide reaction products during prolonged exposure to air \cite{ioni_fe1,ioni_fe2}. 
For an oxide thickness of approximately 1~nm on each surface of the Co and Ni foils, the calculated O K-edge absorbance is only a few times $10^{-3}$ \cite{Henke} and is therefore buried in the noise of the present measurement. 
In contrast, the more extensive oxidation of Fe results in stronger oxygen absorption, allowing the O K-edge feature to be clearly detected.
These results demonstrate that our broadband attosecond SXR platform can simultaneously probe multiple elements, providing access to the L-edges of Ti, Fe, Co, and Ni, as well as the O K-edge, without scanning the photon energy. 
This capability is particularly important for oxygen-containing transition-metal compounds, for which simultaneous access to the transition-metal L-edges and the O K-edge enables future attosecond studies of the coupled dynamics of transition-metal 3d and oxygen 2p electronic states. 
Importantly, our driving laser is wavelength-tunable and energy-scalable, allowing its spectrum to be tailored to maximize the phase-matched photon flux within a desired photon-energy range. 
This scalable approach therefore provides a route to extending the accessible photon-energy range with longer driving wavelengths and to increasing the output through optimized phase matching and a larger interaction cross-sectional area.

\section{Discussion}\label{sec}
\begin{figure}
\centering
\includegraphics[width=\linewidth]{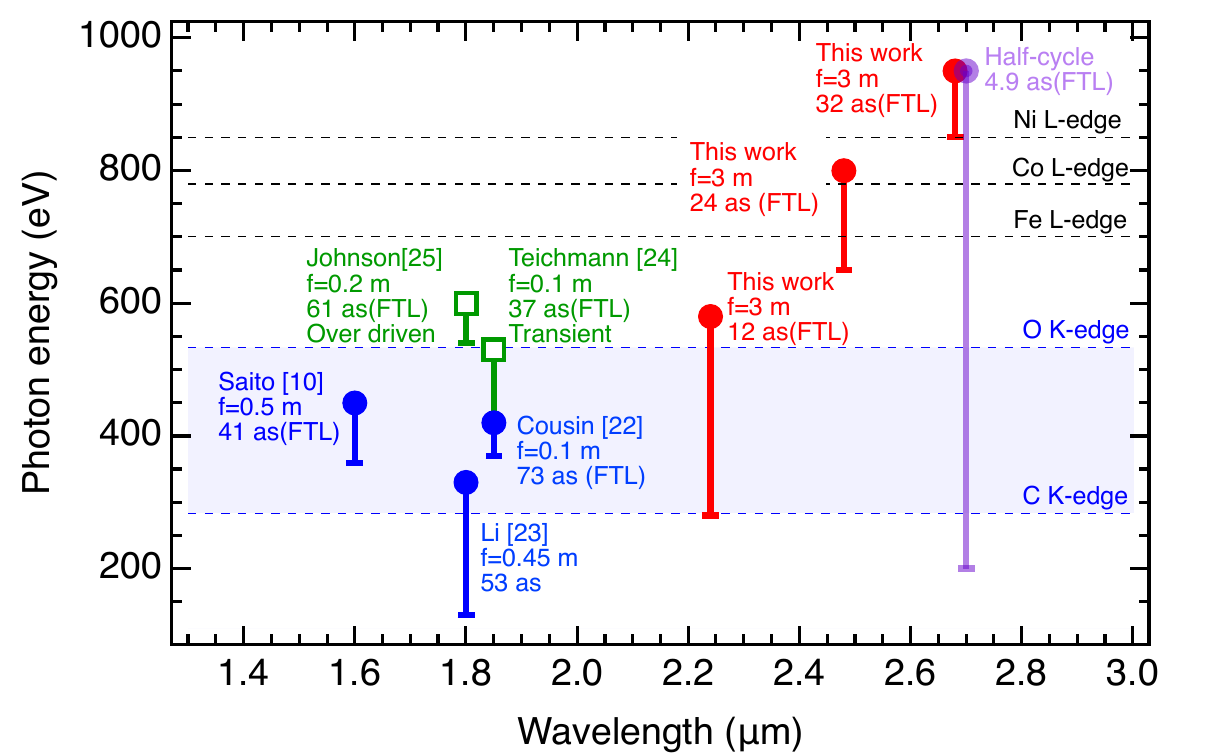}
\caption{
\textbf{Comparison of photon energies achieved by previously reported and present CEP-controlled SXR continua.}\\
Maximum photon energies of CEP-controlled SXR continua generated using different driving wavelengths \cite{bib4,bib17,bib18,bib21,bib22}. 
Symbols indicate the cutoff energies, and vertical bars indicate the measured continuum bandwidths consistent with single-burst emission.
Filled symbols denote CEP-controlled SXR continua generated under phase-matched conditions in neutral gases, whereas open symbols denote those generated using other phase-matching schemes. 
The labels next to each reference indicate the phase-matching method and the FTL pulse duration estimated from the corresponding continuum spectral bandwidth, except for Li et al. \cite{bib22}, for which the experimentally measured pulse duration is given. 
The purple bar indicates the expected SXR continuum accessible with a future 2.7-$\mu$m half-cycle driving laser.
}
\label{fig6}
\end{figure}

To place this work in context, Fig.~\ref{fig6} compares the present source with previously demonstrated attosecond HHG platforms. 
Previous demonstrations of CEP-controlled attosecond pulse generation under phase-matched conditions in neutral gases have been limited to photon energies below approximately 450~eV \cite{bib4}. 
In contrast, IAPs extending to approximately 600~eV have been demonstrated using strongly ionized media and transient phase-matching schemes \cite{bib17,bib18}; however, plasma-induced propagation effects limit coherent buildup over extended interaction lengths and complicate energy scaling. 
The present study bridges this long-standing gap. 
Using phase-matched HHG in a neutral medium, we demonstrate CEP-dependent attosecond emission spanning the entire water-window region and extending to 950~eV in the transition-metal L-edge region. 
This value is approximately twice the maximum photon energy achieved by previous CEP-controlled attosecond sources based on neutral-gas phase matching. 
Furthermore, we experimentally verify the wavelength scaling of the CEP-controlled harmonic cutoff over the 2.2--2.7~$\mu$m driving-wavelength range while maintaining phase-matched operation.

An important remaining challenge is to extend the accessible photon-energy range to higher-energy core-level transitions while maintaining the conditions required for practical attosecond time-resolved spectroscopy.
Following the scaling guideline established for HNC-DCOPA \cite{bib31}, extending the driving wavelength to 3~$\mu$m would increase the cutoff energy to 1.5~keV.
Although the harmonic yield is estimated to decrease by a factor of approximately two, this reduction could be compensated for by increasing the beam radius at the HHG interaction region by a factor of 1.4, thereby enlarging the effective transverse generation area rather than the interaction length.
In the transition-metal L-edge range, the absorption length of He at 1~atm is several tens of centimeters, whereas group-velocity mismatch restricts the effective interaction length to less than 1\% of this value, as discussed in Supplementary Note 2 and reference \cite{GVM}.
In the present setup, the gas-cell length is 50~mm, but the effective interaction length is limited to approximately 7~mm for the $2.68~\mu\mathrm{m}$, three-cycle driving pulse.
Therefore, further scaling of the harmonic yield requires a higher-energy driving laser and a larger beam size at the focus to increase the effective transverse generation area.
Beyond this energy-scaling approach, further shortening the driving pulse toward the half-cycle regime, as expected with HNC-DCOPA \cite{bib31}, would enable substantially broader single-burst SXR emission.
Such a laser would cover the L-edges of multiple ferromagnetic elements and the O K-edge within a continuous spectrum spanning approximately 750~eV, thereby enabling attosecond time-resolved measurements based on single-burst emission (Fig.~\ref{fig6}).
These considerations establish a practical route toward simultaneously scaling the photon energy and photon flux beyond the present transition-metal L-edge range. 
By combining scalable phase matching in a loose-focusing geometry, CEP control, and broadband SXR continua, the present platform offers opportunities for high-sensitivity, element-selective attosecond spectroscopy across an expanded range of transition-metal, correlated, and other functional materials.


\section*{Methods}\label{sec:methods}

\subsection*{Driver laser}\label{subsec:driver-laser}

High-energy mid-infrared driver pulses for HHG were generated using an HNC-DCOPA system \cite{bib31,bib32}. 
Three driver wavelengths were used, centered at 2.26, 2.48, and 2.68~$\mu$m. 
The generation of the 2.26-$\mu$m sub-cycle driver has been described previously \cite{bib31}. 
In the HNC-DCOPA system, the spectral bandwidth was controlled using acousto-optic programmable dispersion filters (AOPDFs). 
The output spectra were tailored using AOPDFs to generate driver pulses centered at 2.48 and 2.68~$\mu$m. 
For operation at 2.68~$\mu$m, the BiB$_3$O$_6$ DCOPA stage for wavelengths below 2.3~$\mu$m was removed, allowing the full pump energy to be transferred to the MgO:LiNbO$_3$ stage that amplifies wavelengths above 2.3~$\mu$m. 
The pulse energies were 28 and 48~mJ for the 2.48- and 2.68-$\mu$m drivers, respectively. 
The pulse durations were characterized by second-harmonic-generation frequency-resolved optical gating (SHG-FROG) using a 10-$\mu$m-thick BBO crystal. 
Pulse durations of 17 and 26~fs were obtained at central wavelengths of 2.48 and 2.68~$\mu$m, corresponding to 1.9 and 2.9 optical cycles, respectively (Figs.~S1 and S2).
\subsection*{HHG setup}\label{subsec:hhg-setup}
High harmonics were generated by focusing the mid-infrared driver pulses into helium gas in a double gas cell \cite{bib40} using a silver-coated concave mirror with a focal length of 3~m.
Helium gas was introduced into the inner gas cell through a gas jet synchronized with the laser pulses, thereby minimizing the gas load on the vacuum chamber.
During operation, the chamber pressure was maintained at approximately 1~Pa.
The helium pressure in the interaction region was estimated to be one-tenth of the backing pressure applied to the gas jet \cite{bib40}.
The generated harmonics were separated from the co-propagating fundamental beam using a 200-nm-thick aluminum filter and analyzed using an SXR spectrometer. 
 For absorption measurements, 200-nm-thick Ti, Fe, Co, or Ni filters were inserted downstream of the aluminum filter.
The spectrometer consisted of a 500-$\mu$m-wide entrance slit, a 2400-groove mm$^{-1}$ diffraction grating (Shimadzu 30-004), and a back-illuminated X-ray CCD detector, providing a spectral resolution of approximately 5~eV.
The pulse energy at the generation point was estimated from the counts recorded by the back-illuminated X-ray CCD detector, accounting for the detector quantum efficiency, the electron yield per photon, the transmission of the Al filter, the diffraction efficiency of the grating, and the spectral fraction selected by the slit.

\subsection*{Phase matching in free-space focusing}\label{subsec}
The optimum phase-matching pressure, $p_{\mathrm{opt}}$, is given by $p_{\mathrm{opt}} = 1/(\pi r_{\mathrm{e}} N_{\mathrm{L}} w_0^2(\eta_{\mathrm{c}}-\eta))$, where $r_{\mathrm{e}}$ is the classical electron radius, $N_{\mathrm{L}}$ is the gas density at 1~atm, $w_0$ is the focal beam radius, $\eta$ is the ionization fraction, $\eta_{\mathrm{c}}$ is the critical ionization fraction given by $2\pi \delta n/(r_{\mathrm{e}} N_{\mathrm{L}} \lambda^2)$, $\lambda$ is the driving-laser wavelength, and $\delta n$ is the refractive-index difference of the neutral gas at 1~atm. 
When the driving intensity is increased such that the ionization fraction exceeds $\eta_{\mathrm{c}}$, the phase-matching condition can no longer be satisfied in a neutral atomic medium; moreover, near this regime, the phase-matching pressure becomes extremely sensitive to small changes in the incident intensity.
Therefore, in this study, we impose the condition $\eta = \eta_{\mathrm{c}}/2$ to represent a weakly ionized neutral-medium regime, in which the ionization fraction varies more gradually with the driving intensity.
This condition determines the maximum allowable peak intensity for a given pulse duration and, once the driving wavelength is specified, determines the corresponding cutoff energy, as shown in Fig.~\ref{fig1}a.
Furthermore, under the condition $\eta = \eta_{\mathrm{c}}/2$, the pressure required for phase matching simplifies to $p_{\mathrm{opt}} = \lambda^2/(\pi^2 \delta n w_0^2)$, indicating that $p_{\mathrm{opt}}$ is proportional to the square of the driving wavelength and inversely proportional to the square of the focal beam radius. 
Thus, the phase-matching pressure can be reduced by increasing the focal beam radius, as shown in Fig.~\ref{fig1}b.

\section*{Supplementary information}

Supplementary information is available for this paper.

\section*{Acknowledgements}
We acknowledge financial support from the Ministry of Education, Culture, Sports, Science and Technology of Japan (MEXT) through the Quantum Leap Flagship Program (Q-LEAP) (grant no. JPMXS0118068681).
K.N. acknowledges the Special Postdoctoral Researcher’s Program of RIKEN and the RIKEN Incentive Research Project.

\section*{Author contributions}
K.N. designed the experimental setup, performed the experiments, and analyzed the experimental data. 
K.N. and E.J.T. discussed the experimental results and wrote the manuscript. 
E.J.T. initiated and supervised this project as a whole.

\section*{Competing interests}

The authors declare no competing interests.

\section*{Data availability}

The data that support the findings of this study are available from the corresponding author upon reasonable request.

\bibliographystyle{apsrev4-2-nodoi}
\bibliography{referenceKeV_v3}

\end{document}